\documentclass[%
reprint,nofootinbib,
 amsmath,amssymb,
 aps,
prl,
]{revtex4-2}

\usepackage{graphicx}% Include figure files
\usepackage{dcolumn}% Align table columns on decimal point
\usepackage{bm}% bold math
\usepackage{braket}
\usepackage{amsmath}
\usepackage{amssymb}
\usepackage{pifont}
\newcommand{\be}{\begin{equation}}
\newcommand{\ee}{\end{equation}}

\begin{document}

\title{The disordered Su-Schrieffer-Heeger model}

\author{Michael Hilke}

\affiliation{%
 Department of Physics, McGill University
}%

%\date{\today}

\begin{abstract}
Quantum topology categorizes physical systems in integer invariants, which are robust to some deformations and certain types of disorder. A prime example is the Su–Schrieffer–Heeger (SSH) model, which has two distinct topological phases, the trivial phase with no edge states and the non-trivial phase with zero-energy edge states. The energy dispersion of the SSH model is dominated by a gap around zero energy, which suppresses the transmission. This exponential suppression of the transmission with system length is determined by the Lyapounov exponent. Here we find an analytical expression of the Lyapounov as a function of energy in the presence of both diagonal and off-diagonal disorder. We obtain this result by finding a recurrence relation for the local density, which can be averaged over different disorder configurations. There is excellent agreement between our analytical expression and the numerical results over a wide range of disorder strengths and disorder types. The real space winding number is evaluated as a function of off-diagonal and on-site disorder for possible applications of quantum topology. 

\end{abstract}

\maketitle 

%\section{Introduction}

The Su-Schrieffer-Heeger (SSH) model represents one of the simplest realizations of topological physics in one dimension \cite{su1979solitons}. Originally proposed to describe the electronic properties of polyacetylene \cite{heeger1988}, the SSH model has gained renewed interest as a theoretical laboratory for understanding topological insulators and their response to disorder \cite{kitaev2006,qi2011}. The model's simplicity—consisting of a one-dimensional chain with alternating hopping amplitudes—makes it an ideal platform for investigating fundamental questions about the interplay between topology and disorder.

In clean systems, the SSH model exhibits a topological phase transition characterized by the opening and closing of an energy gap, accompanied by the appearance or disappearance of topologically protected edge states \cite{zak1982}. The introduction of disorder fundamentally alters this picture, leading to a complex phase diagram where topological protection competes with Anderson localization \cite{anderson1958}. This competition raises profound questions about the nature of topological protection in the presence of disorder and the conditions under which topological edge states can survive \cite{li2009}.

The disordered SSH model has become increasingly relevant due to experimental realizations in various platforms, including  ultracold atoms in optical lattices and self-assembled atoms \cite{atala2013,lohse2016,jalochowski2024implementation}, photonic systems \cite{rechtsman2013,mukherjee2017}, and mechanical metamaterials \cite{wang2015,nash2015}. These implementations have provided direct access to the physics of disordered topological systems, enabling tests of theoretical predictions and the discovery of new phenomena. Interesting works considered disorder in extended SSH models \cite{perez2019interplay,hsu2020topological,cinnirella2024fate}, the SSH model in open systems, \cite{zaimi2021detecting,nava2023lindblad,bissonnette2024boundary} and with correlated disorder or aperiodicity \cite{zuo2022reentrant,liu2022anomalous,sircar2025topological}. 

Beyond its fundamental importance, the disordered SSH model serves as a testing ground for understanding more complex topological systems and their applications in quantum technologies \cite{nayak2008}. The interplay between disorder and topology has implications for quantum computation, where topological protection could provide robust quantum memories, and for understanding the stability of topological phases in realistic materials \cite{alicea2012}.

\begin{figure}[h!]
\includegraphics[width=\linewidth]{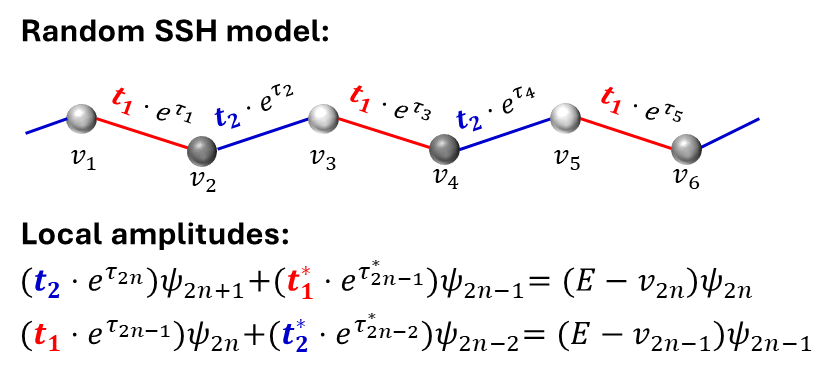}
\caption{\label{SSH} Illustration of the SSH model with alternating hopping $t_1$ and $t_2$, and random corrections $\tau_n$. Further included are random onsite energies $v_n$. The corresponding tight binding equations are shown for the local amplitudes $\psi_n$.
}
\end{figure}

The disordered SSH model is illustrated in figure \ref{SSH}, which generalizes the SSH model by including random hopping variables, $\tau_i$ and adding random onsite potentials, $v_i$. Taking $\tau_i=0$ and $v_i=0$ reproduces the conventional SSH model.

An interesting simpler case that was examined early on is the non-topological case, where $t_1=t_2$ in the presence of off-diagonal disorder. In this case, anomalous behavior emerges near zero energy as a consequence of chiral symmetry. Under a mapping $\psi_n \rightarrow (-1)^n \psi_n$, the Hamiltonian changes sign. As a result, the eigenvalues occur in pairs $\pm E$. At zero energy, this additional sublattice symmetry affects the localization properties of the off-diagonal disorder case. Indeed, at $E=0$ the localization length diverges \cite{theodorou1976extended} and the envelope decays as $\sim e^{-\lambda_2\sqrt{N}}$ \cite{soukoulis1981off,ziman1982localization}. This anomalous behavior is related to the bipartite sublattice structure that naturally emerges at $E=0$ \cite{inui1994unusual}. By contrast, in the SSH model, the bipartite sublattice structure is intrinsic and persists at all energies, leading to qualitatively different localization properties.

To solve the disordered SSH model, we start with the generalized tight-binding equation for real energy $E$ and real onsite potentials $v_n$ ($E_n\equiv E-v_n$) and hopping $T_n\equiv t_\epsilon e^{\tau_n}$, where $\epsilon=1$ or 2, for $n$ odd and even, respectively.
\be 
T_n\psi_{n+1}+T_{n-1}^*\psi_{n-1}=E_n\psi_n
\label{psi}
\ee
for the amplitudes $\psi_n$ (which can be complex). The solution can be computed numerically very efficiently by solving the corresponding iterative transfer matrix equation
\be \begin{pmatrix}
\psi_{n+1}\\ \psi_n\end{pmatrix}=\begin{pmatrix}E_n/T_n & -T_{n-1}^*/T_n\\1 & 0\end{pmatrix}\begin{pmatrix}
\psi_{n} \\ \psi_{n-1}\end{pmatrix},
\ee
where the Lyapounov exponent can be computed as 
\be \lambda_N=\frac{2}{N}
\log \left( \max \left[ eig\prod_{n=1}^N \begin{pmatrix}E_n/T_n & -T_{n-1}^*/T_n\\1 & 0\end{pmatrix}\right]\right),
\label{lyapounovnum}
\ee
where we take $T_0=0$ and $N\rightarrow\infty$. The factor 2 is a result of squaring the amplitudes for the transmission. Taking $T_0=0$ is equivalent to defining a semi-infinite chain, where $\psi_n=0$ for $n\leq 0$. This also fixes the gauge of the SSH model, i.e., the sublattice structure. We will always consider the case $N$ even, so that no unit cell is cut. To compute $\lambda$ numerically we use expression \eqref{lyapounovnum}. The associated localization length is simply the inverse of the Lyapounov exponent. 

The main goal here is to derive an analytical expression of the Lyapounov exponent. Our approach is quite different from the work in ref. \cite{perez2019interplay} as it is non-perturbative. In order to achieve this, and since 
the transmission depends on the squared amplitudes $\rho_n\equiv|\psi_n|^2$ (local density) it is useful to
write the corresponding iterative equation for the local densities, which yields
\be 
\rho_{n+1} = \sum_{i=0}^2 D_n^{(i)}\rho_{n-i}
\label{rho}
\ee
with
\begin{align*}
D^{(0)}_n &=\left[\frac{E_n^2}{|T_n|^2}-\frac{|T_{n-1}|^2 E_n}{|T_n|^2 E_{n-1}}\right]\\
D^{(1)}_{n}&=
\left[\frac{|T_{n-1}|^2}{|T_n|^2}-\frac{E_n E_{n-1}}{|T_n|^2}\right]\\
D^{(2)}_n&=\left[\frac{|T_{n-2}|^2 E_n}{|T_n|^2 E_{n-1}}\right].
\end{align*}
This equation only depends on real quantities and can be obtained from eq. \eqref{psi} using simple algebra (see supplemental material).

The analytical expression for $\lambda$ is obtained by disorder averaging \eqref{rho}, which yields (see the supplementary material).
\be 
\bar{\rho}_{n+1}= \sum_{i=0}^2\bar{D}^{(i)}_\epsilon\bar{\rho}_{n-i}
\label{avrho}
\ee
with
\begin{align*}
\bar{D}^{(0)}_\epsilon &=\overline{T_\epsilon^{-2}}\left[E^2+\overline{v_\epsilon^2}-\left(\overline{{T_{\bar{\epsilon}}^{-2}}}\right)^{-1} \right]\\
\bar{D}^{(1)}_\epsilon&=\overline{T_\epsilon^{-2}}\left[-E^2+\overline{v_{\bar{\epsilon}}^2}+\left(\overline{{T_{\bar{\epsilon}}^{-2}}}\right)^{-1} \right]\\
\bar{D}^{(2)}_\epsilon&=\overline{T_\epsilon^{-2}}\,\cdot\,\overline{T_\epsilon^2}
\end{align*}

For uniformly distributed random variables $|v_n|<V_\epsilon/2$ and $|\tau_n|<W_\epsilon/2$, we have $\overline{v_\epsilon^2}=V_\epsilon^2/12$, $\overline{T_\epsilon}=t_\epsilon\, \text{sh}(W_\epsilon/2)$, $\overline{T_\epsilon^2}=t_\epsilon^2\,\text{sh}(W_\epsilon)$, and $\overline{T_\epsilon^{-2}}=t_\epsilon^{-2}\,\text{sh}{(W_\epsilon)}$, for $n$ either even ($\epsilon$) or odd ($\bar{\epsilon}$). For convenience, we defined $\text{sh}(x)\equiv \sinh{x}/x$. The disorder is assumed to be uncorrelated: $\langle v_n v_m\rangle=(V_\epsilon^2/12)\delta_{n,m}$ and $\langle \tau_n \tau_m\rangle=(W_{\epsilon}^2/12)\delta_{n,m}$.
The hopping elements $t_\epsilon$ alternate between values $t_1$ and $t_2$. For $t_1=t_2=1$ we recover the random Anderson model, which leads to localization for uncorrelated disorder. Expression \eqref{avrho} is consistent with the disordered Anderson model without off-diagonal disorder \cite{hilke2008ensemble} and represents a generalization of this earlier result by including off-diagonal disorder and alternating hopping as relevant for the SSH model. Without disorder we recover the conventional SSH model for $t_1\neq t_2$.

\begin{figure}[h!]
\includegraphics[width=\linewidth]{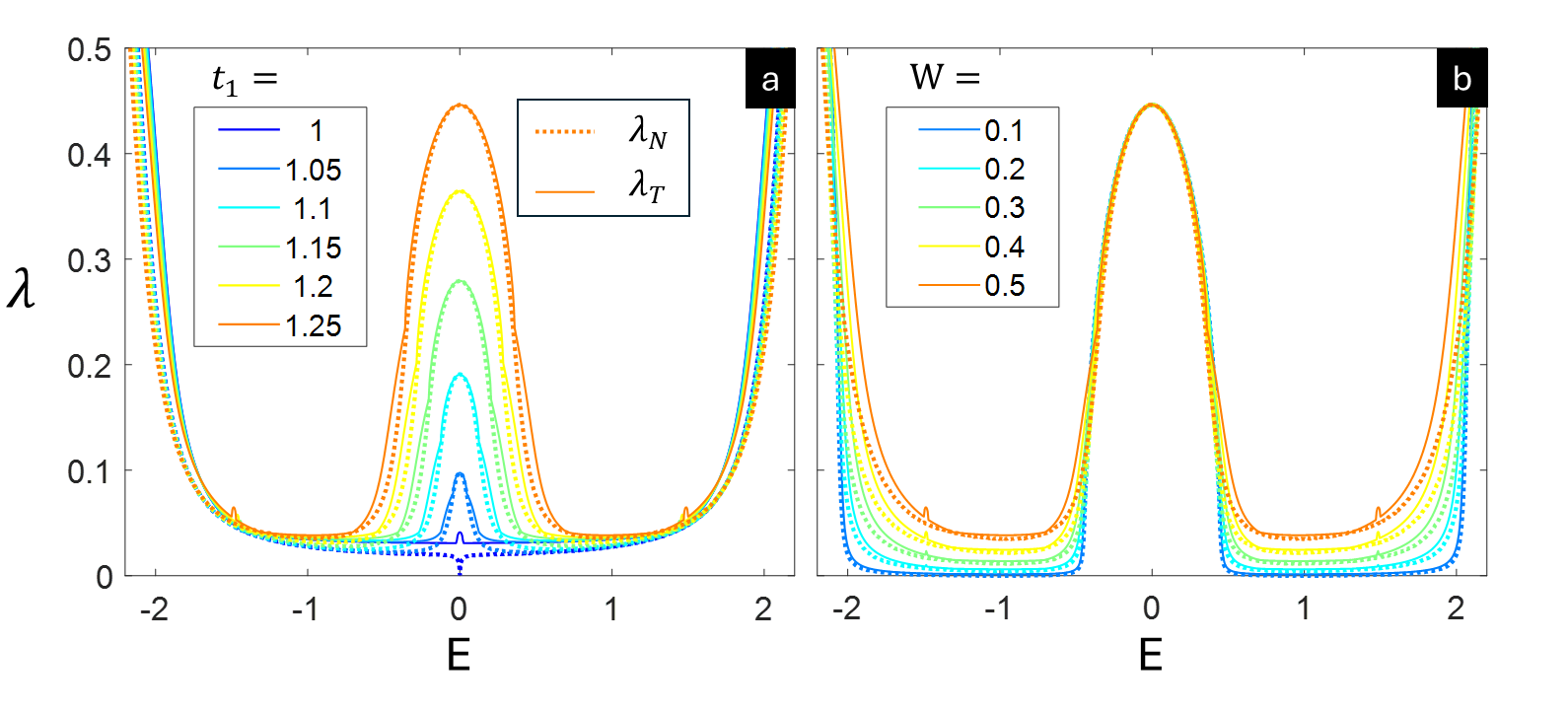}
\caption{\label{lyapounov} Lyapounov exponent as a function of energy. In (a) the plot shows different values of $t_1$ from 1 to 1.25 with $t_2=1/t_1$ and with off-diagonal disorder with $W=0.5$. In (b) different values of off-diagonal disorder strengths are shown for $W=0.1$ to $0.5$ and $t_1=1.25$ and $t_2=1/t_1$. The Lyapounov exponent is computed using $\lambda_N$ from equ. \eqref{lyapounovnum}, where we used $N=2000$ and 1000 configurational averages. For $\lambda_T$ we used equ. \eqref{lypounovT}.
}
\end{figure}

We are now in the position to compute the localization properties of the disordered SSH model. For simplicity, we will consider the case of uniform disorder (not alternating between even and odd sites), i.e.,  $W_\epsilon=W_{\bar{\epsilon}}=W$. We will start with the non-topological phase, where $t_1>t_2>0$. In this case the conventional SSH model has a gap at the band center and no topological edge states at zero energy. To compute the Lyapounov exponent numerically we use equ. \eqref{lyapounovnum} for a finite chain length $N$ and show the results in FIG. \ref{lyapounov}. 

To compute the analytical expression for the SSH model we need to evaluate the eigenvalues of 
\be
\Lambda=\begin{pmatrix}
\bar{D}^{(0)}_1 & \bar{D}^{(1)}_1 & \bar{D}^{(2)}_1 \\
1 & 0 & 0\\
0 & 1 & 0\end{pmatrix}
\begin{pmatrix}
\bar{D}^{(0)}_2 & \bar{D}^{(1)}_2 & \bar{D}^{(2)}_2 \\
1 & 0 & 0\\
0 & 1 & 0\end{pmatrix}
\label{L}. 
\ee
The indices 1 and 2 represent the even and odd solutions in equation \eqref{avrho}. Unlike the transfer matrices in equation \eqref{lyapounovnum}, where there are only two eigenvalues, one representing the decaying solution and the other the diverging solution, here we have 3 eigenvalues of $\Lambda$. The Lyapounov exponent can then be obtained from 
\be 
\lambda_{T}=\text{Max}_2|\log |\text{eig}(\Lambda)||/2
\label{lypounovT}
\ee
The factor of 2 comes from considering the transfer matrix corresponding to the unit cell of 2 sites and $\text{Max}_2$ is defined as the average of the two largest eigenvalues, which gives the representative average divergence. The agreement between the analytical Lyapounov exponent, $\lambda_{T}$, and the numerical one ($\lambda_{N}$) is very good as shown in FIG. \ref{lyapounov}. At zero energy ($E=0$) the Lyapounov exponent simplifies to $\lambda=\log(t_1/t_2)$ consistent with the numerical value and the presence of off-diagonal disorder does not affect $\lambda$. At $E=0$ off-diagonal disorder does not break the chiral symmetry of the SSH model and doesn't affect the decay of the zero energy state. Without disorder, the periodic SSH model (infinite length) is gapped for $|E|<|t_1-t_2|$, which implies that $\lambda>0$ inside the gap in the absence of disorder. This confirms that in the gap there are no extended states in the periodic SSH model. Outside the gap, $|t_1-t_2|<|E|<|t_1+t_2|$ and without disorder $\lambda=0$, while with disorder $\lambda>0$ and all states are Anderson localized. 

We hereby turn to the topological phase, i.e., $0<t_1<t_2$. The topology of the SSH model is best described by the Zak phase, which 
is defined as the geometric phase acquired by the Bloch wavefunction as it traverses the entire Brillouin zone. It is equivalent to the Berry phase in 1D systems across a non-contractible loop \cite{zak1989berry}. The Zak phase can be related to the polarization \cite{benalcazar2017quantized} and winding number \cite{asboth2016}. The Zak phase is defined as
$Z = \int_{-\pi}^{\pi} A(k) dk$, 
where $A(k) = i \braket{u(k) | \partial_k | u(k)}$ represents the  Berry connection of the Bloch eigenstate $\ket{u(k)} $. The Zak phase is defined for an infinite periodic system, where Bloch's theorem can be used. For a finite system ($n=1$ to $N$) and $N$ even, we can use the bulk-boundary correspondence, to relate the winding number ($\gamma=Z/2\pi$) to the number of topological edge states ($=2\gamma$) inside the gap. For $t_1<t_2$ the winding number is $\gamma=1$ and there are two edge states inside the gap (close to $E=0$) \cite{asboth2016}. For $t_1>t_2$, the winding number is 0 and no edge state. The factor 2 in the number of edge states for $N$ even is due to chiral symmetry, which forces the spectrum to be symmetric in energy ($\pm E$) and the edge states occur in pairs. By comparison, the extended SSH model, where long-range hopping is included, can have any integer number of pairs of edge states close to zero energy and a winding number that is an arbitrarily large integer (both positive and negative) \cite{MOOLA2026131272}. 

\begin{figure}[h!]
\includegraphics[width=\linewidth]{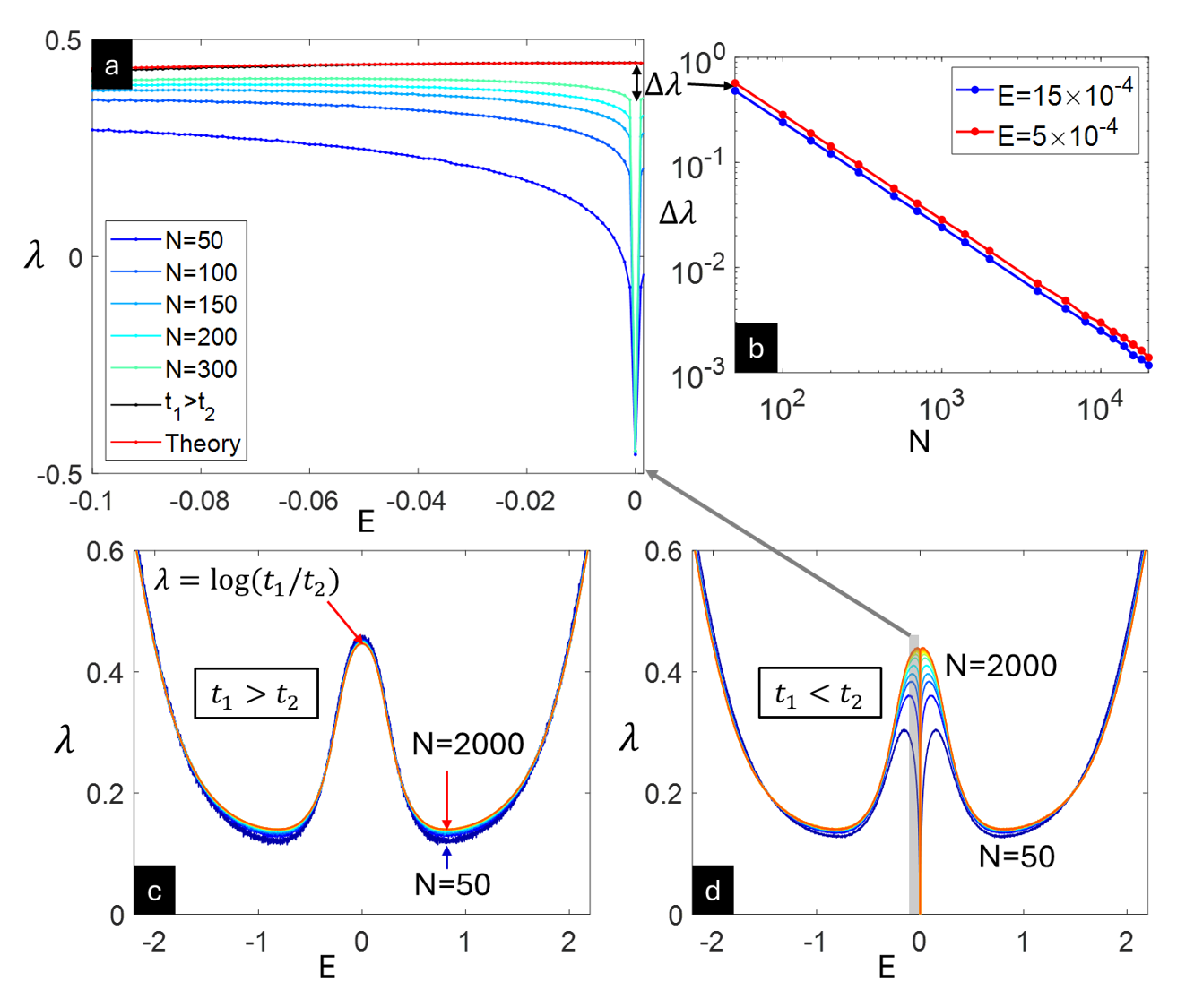}
\caption{\label{TopononTopo} In (a) we show the topological $\lambda_N$ as a function of energy for different values of $N$ for $t_1=0.8$ and $t_2=1/t_1$. Also shown is the non-topological $\lambda_N$ for $N=2000$ and $t_1=1/0.8$ and $t_2=1/t_1$ as well as $\lambda_T$ for the same parameters. In b) $\Delta\lambda=
|\lambda_N(E)-|\log(t_1/t_2)||$ is shown as a function of $N$ for $t_1=0.8$ and $t_2=1/t_1$ for 2 values of $E$. In c) and d) $\lambda_N$ is shown as a function of $E$ for different values of $N$ for the non-topological and topological cases, respectively. We used ($t_1=1/0.8$, $t_2=1/t_1$) and ($t_1=1/0.8$, $t_2=1/t_1$), respectively. 
}
\end{figure}

We now compare the topological ($t_1<t_2$, FIG. \ref{TopononTopo}d) case to the non-topological ($t_1>t_2$, FIG. \ref{TopononTopo}c) case with disorder. The main difference between the two cases is seen close to zero energy because of the existence of topological edge states for $t_1<t_2$. These edge states manifest themselves through a change of sign of $\lambda$ close to zero because of the bipartite nature of the topological edge state at $E=0$. Indeed, the amplitude is zero  ($\psi_n=0$) for $n$ even and decays as $|\psi_n|\sim (t_1/t_2)^{(n-1)/2}$ for $n$ odd, which gives rise to the robust negative Lyapounov exponent of $\lambda=\log(t_1/t_2)$ at $E=0$ (see FIG. \ref{TopononTopo}a). For $t_1>t_2$ there are no solutions in the gap and the Lyapounov exponent as defined in equ. \eqref{lyapounovnum} is positive, which illustrates that for a fixed amplitude at $n=1$ the solution would explode with $N$. This exponential divergence has the same absolute rate as the decay rate of the topological edge state at exactly $E=0$. Slightly away from zero energy ($E=0+\epsilon$) $\lambda$ depends strongly on $N$ and the difference vanishes as $\Delta\lambda=
|\lambda_N(\epsilon)-|\log(t_1/t_2)||\sim 1/N$ as illustrated in FIG. \ref{TopononTopo}b. Unlike what was found in \cite{perez2019interplay}, we find that in the limit of $N\rightarrow \infty$ there is no difference between the Lyapounov exponent in the topological vs. non-topological case for $E\neq 0$ and at $E=0$ the absolute values of the Lyapounov exponent are identical.

\begin{figure}[h!]
\includegraphics[width=\linewidth]{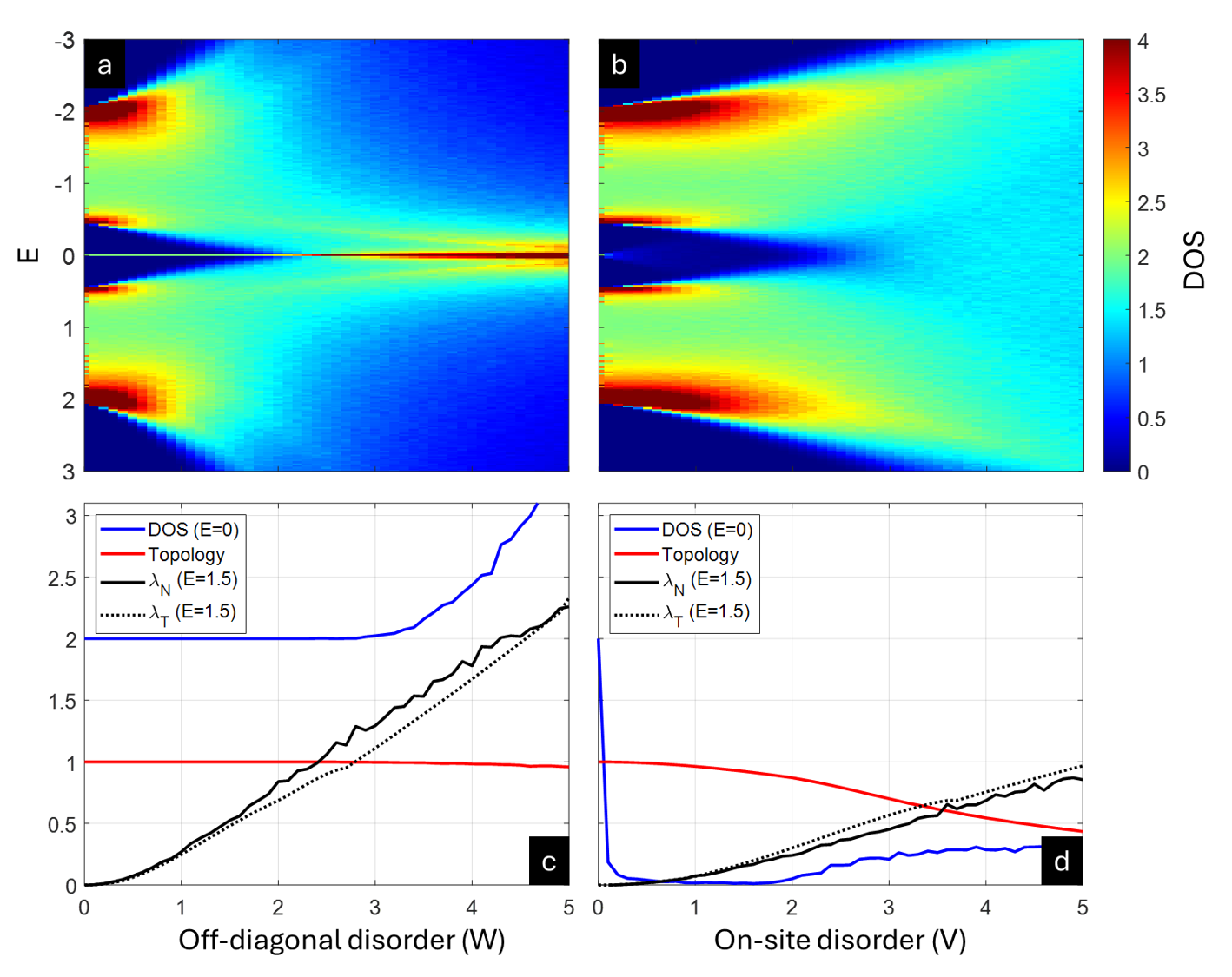}
\caption{\label{DOS} The (a) and (b) panels show the density of states as a function of off-diagonal ($W$) (a) and diagonal ($V$) (b) disorder strength. The (c) and (d) panels show the DOS at $E=0$, the real space winding number, and the Lypounov exponent (the theoretical and numerical value) as a function of off-diagonal (a) and diagonal (b) disorder strength. The red line shows the winding number as a function of the disorder strength. Here $t_1=0.8$ and $t_2=1/t_1$.
}
\end{figure}

To further characterize the effect of disorder on the topological phase of the SSH model, it is instructive to compute the density of states (DOS) of a finite chain of length $N$ ($N$ is even). The DOS is obtained by counting the number of eigenvalues within an energy interval $\Delta E=4/N$, which defines a DOS largely independent of $N$ and is plotted in FIG. \ref{DOS}. For $t_1<t_2$, which is the topological phase of the SSH model, this leads to the existence of 2 zero energy states, which are seen at the center of the band in the absence of disorder. With increasing disorder the DOS at $E=0$ stays constant for the off-diagonal disorder at small enough disorder, which illustrates the topological protection of the $E=0$ state with respect to off-diagonal disorder, while the DOS of the $E=0$ state rapidly vanishes in the presence of on-site disorder, which is not protected by the chiral symmetry of the SSH model. Eventually the gap closes. Similar behavior can also be seen for generalized SSH models, such as SSH$_4$ and the extended SSH model \cite{MOOLA2026131272}.

It is possible to characterize the topology as a function of disorder, which can be obtained by computing the local real space winding number following ref. \cite{oliveira2024robustness}
\be
\gamma_n=4\,\text{Diag}(W\rho_+X\rho_-),
\ee
where $W$ is the chiral operator (alternating 1 and -1 on the diagonal), $X$ the position operator $(1\,1\,2\,2\,\cdots N/2\,N/2)$ on the diagonal, and $\rho_\sigma$ the density operator of occupied levels ($\sigma=-$ for $E<0$) and unoccupied levels ($\sigma=+$ for $E>0$). The bulk winding number is obtained by averaging $\gamma_n$ inside the chain by avoiding the edges ($\gamma\simeq\langle \gamma_n \rangle_{N-edge}$). $\gamma$ is no longer perfectly quantized in the presence of disorder, but stays very close to 1 for off-diagonal disorder while quickly moves away from 1 for on-site disorder.  

\begin{figure}[h!]
\includegraphics[width=\linewidth]{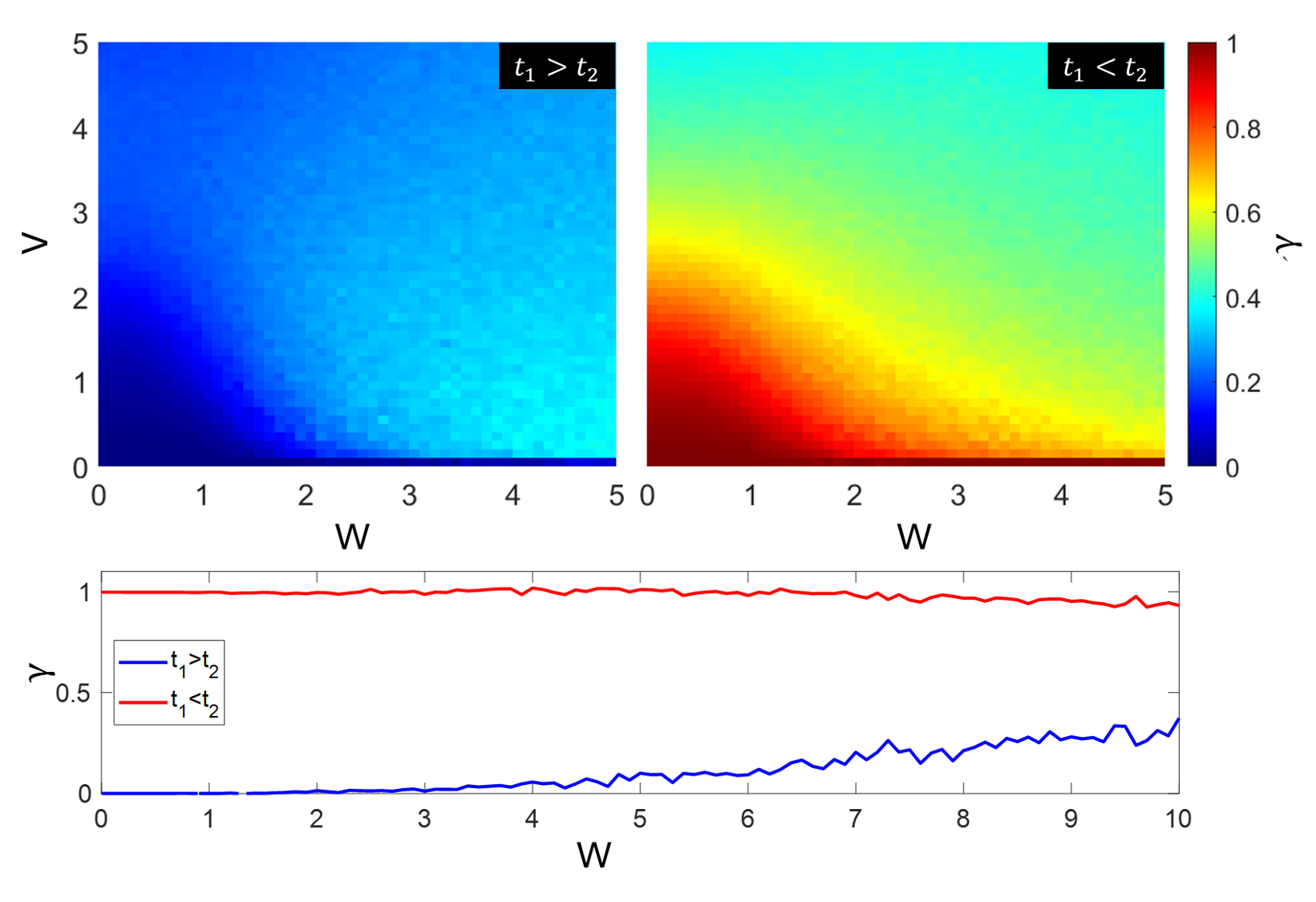}
\caption{\label{winding} The top panels show the real space winding number $\gamma$ (color scale) as a function of off-diagonal disorder ($W$) and on-site disorder ($V$) for $t_1=0.9$ (left) and $t_1=1/0.9$ (right), respectively. In both cases $t_2=1/t_1$. The bottom plot depicts the two cases for $V=0$.
}
\end{figure}

A more complete picture of the real space winding number as a function of off-diagonal disorder ($W$) and on-site disorder ($V$) is shown in FIG. \ref{winding}. The figure illustrates the fundamental difference between on-site disorder and off-diagonal disorder. For zero off-diagonal disorder the real space winding number depends only weakly on the disorder strength, illustrating the topological robustness to off-diagonal disorder over a wide range. This is arguably the most important hallmark of topology for applications of quantum topology. 

Summarizing, we have derived an analytical expression for the Lyapounov exponent to determine the localization properties of the random SSH model. An excellent agreement was found between the analytical expression and the numerical results. We could demonstrate that in the limit of large length, the topological phase does not affect the localization length. We further characterized the topology of the random SSH model as a function of off-diagonal and on-site disorder, Our results can be extended to generalized SSH models with more than two atoms per unit cell.   

The author acknowledges support from NSERC, INTRIQ and RQMP.

\bibliography{refs}% Produces the bibliography via BibTeX.

\begin{widetext}
\section{Supplementary material}

\subsection{Derivation of equation \eqref{rho}}
To derive the equation for the local densities, we start with \eqref{psi}, where we can write 
\be
\psi_{n+1}=\frac{E_n}{T_n}\psi_n-\frac{T^*_{n-1}}{T_n}\psi_{n-1}.
\ee
Writing this equation in terms of the local density, we have
\be
\rho_{n+1}\equiv\psi_{n+1}\psi^*_{n+1}=\frac{E_n^2}{|T_n|^2}\rho_n+\frac{|T_{n-1}|^2}{|T_n|^2}\rho_{n-1}-\frac{E_n(T^*_{n-1}\psi_n^*\psi_{n-1}+h.c.)}{|T_n|^2}
\ee
and doing the same for $\psi_{n-2}$, we have 
\be
\rho_{n-2}=\frac{E_{n-1}^2}{|T_{n-2}|^2}\rho_{n-1}+\frac{|T_{n-1}|^2}{|T_{n-2}|^2}\rho_{n}-\frac{E_{n-1}(T^*_{n-1}\psi_n^*\psi_{n-1}+h.c.)}{|T_{n-2}|^2}.
\ee
Combining these two equations allows us to eliminate the term $(T^*_{n-1}\psi_n^*\psi_{n-1}+h.c.)$, which leads to equation \eqref{rho}, where $h.c.$ is the Hermitian conjugate.

\subsection{Derivation of equation \eqref{avrho}}

To derive the iterative equation for the averages, we start with equation 
\eqref{rho}, which can be written as
\be 
\rho_{n+1} = \frac{E_n^2}{|T_n|^2}\rho_n-\frac{|T_{n-1}|^2 E_n}{|T_n|^2 E_{n-1}}\rho_n+ D_n^{(1)}\rho_{n-1}+D_n^{(2)}\rho_{n-2}.
\label{rho1}
\ee
Since $\rho_{n}$ depends on $T_{n-1}$ and $E_{n-1}$ we cannot directly take the disorder average and separate the averages of $\rho_n$ and its coefficients. We therefore have to rewrite equation \eqref{rho1} by eliminating the coefficients that cannot be separated by averaging. We do this in two steps. The first step allows us to separate $E_{n-1}$ from $\rho_n$ and in the second step we will separate $T_{n-1}$ from $\rho_n$. Using the iterative equation \eqref{rho1} to compute the previous element $\rho_n$, we can write
\be
\frac{E_n}{E_{n-1}}\rho_n=\rho_n+\left(\frac{E_n}{E_{n-1}}-1\right) \sum_{j=0}^2 D_{n-1}^{(j)}\rho_{n-1-j}
\ee
Plugging this back into \eqref{rho1}, yields
\be 
\rho_{n+1} = \frac{E_n^2}{|T_n|^2}\rho_n-\frac{|T_{n-1}|^2}{|T_n|^2}\rho_n-\frac{|T_{n-1}|^2}{|T_n|^2}\left(\frac{E_n}{E_{n-1}}-1\right)\sum_{j=0}^2 D_{n-1}^{(j)}\rho_{n-1-j}+D_n^{(1)}\rho_{n-1}+D_n^{(2)}\rho_{n-2}
\equiv\sum_{j=0}^3 F_n^{(j)}\rho_{n-j},
\label{rhoE}
\ee
where have defined new coefficients $F_n^{(j)}$. Note that we have eliminated the factor $E_{n-1}$ multiplying $\rho_n$. We still have coefficient $T_{n-1}$, which we proceed to eliminate next. For this step we introduce $\alpha_\epsilon$ and use equation \eqref{rhoE} for $n-1$ to write
\be 
\frac{|T_{n-1}|^2}{|T_n|^2}\rho_n=\alpha_\epsilon\rho_n+\left(\frac{|T_{n-1}|^2}{|T_n|^2}-\alpha_\epsilon\right)\sum_{j=0}^3 F_{n-1}^{(j)}\rho_{n-1-j},
\ee
which we insert into \eqref{rhoE} to yield
\be
\begin{aligned}
\rho_{n+1}& =\left( \frac{E_n^2}{|T_n|^2}-\alpha_\epsilon\right)\rho_n-\left(\frac{|T_{n-1}|^2}{|T_n|^2}-\alpha_\epsilon\right)\sum_{j=0}^3 F_{n-1}^{(j)}\rho_{n-1-j}-\frac{|T_{n-1}|^2}{|T_n|^2}\left(\frac{E_n}{E_{n-1}}-1\right)\sum_{j=0}^2 D_{n-1}^{(j)}\rho_{n-1-j}+D_n^{(1)}\rho_{n-1}+D_n^{(2)}\rho_{n-2}
\\
&\equiv\sum_{j=0}^4 G_n^{(j)}\rho_{n-j},
\end{aligned}
\label{rhoG}
\ee
Using the definitions of the coefficients $D^{(j)}_n$ and $F^{(j)}_n$ and after some algebra, we obtain
\begin{align*}
G^{(0)}_n &=\left[\frac{E_n^2}{|T_n|^2}-\alpha_\epsilon\right]\\
G^{(1)}_{n}&=
\left[\frac{|T_{n-1}|^2}{|T_n|^2}-\frac{2E_n E_{n-1}-E_{n-1}^2}{|T_n|^2}+\frac{|T_{n-2}|^2(E_n-E_{n-1})}{|T_n|^2E_{n-2}}+(|T_{n-2}|^2-E^2_{n-1})\left(\frac{1}{|T_n|^2}-\frac{\alpha_\epsilon}{|T_{n-1}|^2}\right)\right]\\
G^{(2)}_n&=\cdots.
\end{align*}
Here we only show the averaging procedure for $G^{(1)}_{n}$ since the other coefficients involve very similar procedure. 
These coefficients simplify greatly when we take the disorder average. We illustrate this with the expression $G^{(1)}_{n}$, which contains terms $T_{n-2}$ and $E_{n-2}$, which would appear in front of $\rho_{n-1}$. Luckily, by taking the disorder average, the term $(E_n-E_{n-1})$ can be separated out and is zero by evaluating the disorder average $\langle E_n\rangle=E$. The same is true for the term $\left(\frac{1}{|T_n|^2}-\frac{\alpha_\epsilon}{|T_{n-1}|^2}\right)$, which can be separated out and averages out to zero if we take $\alpha_\epsilon\equiv\langle 1/|T_n|^2\rangle/\langle 1/|T_{n-1}|^2\rangle$. Note that in the SSH model $\alpha_\epsilon$ can take on two possible values depending on whether $n$ is odd or even. Taking the disorder average of all the coefficients yields the result given in equation \eqref{avrho}, i.e., $\langle G^{(j)}_n \rangle=D^{(j)}_\epsilon$ and $\langle G^{(3)}_n \rangle=\langle G^{(4)}_n \rangle=0$, where $\epsilon$ indicates the even or odd case.

\end{widetext}
\end{document}